\documentclass[aps,prl,10pt,superscriptaddress,showpacs,twocolumn, longbibliography]{revtex4-2}

\usepackage[dvipdfmx]{graphicx}
\usepackage{amsfonts}
\usepackage{dsfont}
\usepackage{mathtools}
\usepackage{amssymb}
\usepackage{csquotes}
\usepackage{xcolor}
\usepackage{orcidlink}
\usepackage{tikz-cd}
\usepackage{enumitem}
\usepackage{hyperref}
\usepackage{dcolumn}
\usepackage{bm}
\usepackage{amsmath}

\usepackage[capitalise,nameinlink]{cleveref}
\usepackage[authormarkuptext=name,commentmarkup=uwave]{changes}
\definechangesauthor[name={JRH},color=green!70!black]{jonte}

\usepackage[normalem]{ulem}

\begin{document}


\title{Quantum Contextuality Requires Counterfactual Gain}

\author{Yuki Sagawa}
\affiliation{Program in Advanced Materials Science, Faculty of Engineering and Design, Kagawa University, 2217-20 Hayashi-cho, Takamatsu, Kagawa 761-0396, Japan}

\author{Jonte R. Hance}
\email{Jonte.Hance@newcastle.ac.uk}
\affiliation{School of Computing, Newcastle University, 1 Science Square, Newcastle upon Tyne, NE4 5TG, UK}
\affiliation{Quantum Engineering Technology Laboratories, Department of Electrical and Electronic Engineering, University of Bristol, Woodland Road, Bristol BS8 1US, UK}

\author{Holger F. Hofmann}
\email{hofmann@hiroshima-u.ac.jp}
\affiliation{Department of Quantum Matter, Graduate School of Advanced Science and Engineering, Hiroshima University, Kagamiyama 1-3-1, Higashi Hiroshima 739-8530, Japan}

\author{Takafumi Ono}
\email{ono.takafumi@kagawa-u.ac.jp}
\affiliation{Program in Advanced Materials Science, Faculty of Engineering and Design, Kagawa University, 2217-20 Hayashi-cho, Takamatsu, Kagawa 761-0396, Japan}


\date{\today}

\begin{abstract}
Quantum contextuality, where measurement outcomes depend on the measurement context, implies a failure of classical realism in quantum systems. As recently shown, the transition between measurement contexts can be mapped onto the path that a quantum particle takes through an interferometer. Here, we investigate the relation between contextuality and the counterfactual gain observed in the output ports of such an interferometer when one of the paths is blocked. It is shown that experimental evidence of contextuality can only be obtained when counterfactual gain is observed for a specific combination of blocked path and output port. Using a silicon photonic integrated circuit, we experimentally observe the counterfactual gain for a selection of input states and evaluate the associated evidence for contextuality. The results confirm that contextuality can only be observed in the presence of counterfactual gain. 

\end{abstract}

\maketitle
\section{Introduction}
Quantum mechanics exhibits many phenomena that cannot be explained by classical physics. One of the most fundamental of these phenomena is quantum contextuality, where measurement outcomes cannot be consistently explained by assuming the existence of pre-established, non-contextual properties that are independent of the choice of measurement \cite{KochenSpecker1967, Budroni2022}. This behaviour fundamentally contradicts the classical notion of objective realism, where measurement outcomes are assumed to exist independently of the measurement context.

The existence of quantum contextuality was first systematically demonstrated by Kochen and Specker \cite{KochenSpecker1967}. They showed, through a complex construction involving more than one hundred vectors, that assuming a noncontextual hidden variable model in a Hilbert space of dimension $\geq3$ leads to a logical contradiction. Subsequently, more concise proofs have been proposed, such as Hardy’s arguments based on paradoxes \cite{Hardy1992,Hardy1993}, the KCBS inequality for spin-1 systems introduced by Klyachko and collaborators \cite{Klyachko2008}, and the minimal Kochen–Specker-type proof presented by Cabello and colleagues \cite{Cabello2013}. Furthermore, Spekkens extended the notion of contextuality to a more general operational framework, constructing a theory which (it is claimed) encompasses a broader range of physical operations \cite{Spekkens2005} (though its usefulness as a measure of the ``quantumness'' of a system has been challenged \cite{tezzin2025ontologicalmodelsadequatelyrepresent}).

In addition to these theoretical developments, contextuality has been experimentally tested across a variety of physical platforms. For example, Bell–Kochen–Specker-type experiments have been conducted using photons \cite{Huang2003,DAmbrosio2013,Zhang2019,Qu2021,Liu2023}, trapped ions \cite{Kirchmair2009,Zhang2013,Leupold2018,Wang2022}, superconducting qubits \cite{Jerger2016}, and solid-state spin systems \cite{George2013}. In many of these experiments, contextuality has been tested by switching between different measurement settings and comparing the results. However, such comparisons between separately obtained measurement outcomes make it difficult to directly observe the relationships between different measurement contexts, or the internal dynamics of the system.


This led Hofmann to propose a method for testing contextuality using a three-path interferometer, which allows us to investigate the dynamics of photon propagation within this system \cite{Hofmann2023}. The interferometer consists of three input modes, three output modes, and five beam splitters that implement the transitions between five different measurement contexts---all through the propagation of single photons. Furthermore, he proposed a method for analysing the conditional currents inside the interferometer via weak measurements, allowing for the direct verification of the existence of negative conditional currents arising from quantum contextuality.

Hance \textit{et al} showed that these same negative conditional currents can also provide a counterfactual gain in our ability to detect the presence of an absorber even when no photon is absorbed \cite{Elitzur1993Bomb,Hance2024}. 
This counterfactual gain refers to the gain in our ability to discriminate whether or not an absorber has been placed on a certain path, originating from cases where the change in the output probability distribution caused by inserting the absorber exceeds what would be expected from mere photon loss, leading to results that contradict classical predictions.
In Hofmann's contextual interferometer, a negative conditional current between a path inside the interferometer and an output port predicts a counterfactual gain towards that output port when that path inside the interferometer is blocked. 
It should therefore be possible to use counterfactual gain as a method of characterising the nonclassical statistics of the interferometer responsible for the violation of noncontextual inequalities. 

In this paper, we identify the relation between the existence of a counterfactual gain for outputs of an interferometer, and the ability to violate a noncontextual inequality for the probabilities of photon detection on paths inside the interferometer. 
We show that the noncontextual inequality can only be violated by interferometers where we could observe a counterfactual gain for a specific combination of blocked path and output port. 
We experimentally demonstrated a violation of this condition by up to 21 \%, employing a silicon photonic integrated circuit. 
Our results reveal a fundamental relation between counterfactual gain and contextuality, motivating further research into both the role and the manifestations of contextuality in optical quantum circuits. This should offer significant insights into both the foundations of quantum mechanics and its applications in quantum information science.

\section{Quantum Contextuality in Hofmann's Three-Path Interferometer}
\begin{figure}[t]
 \centering
 \includegraphics[width=\columnwidth]{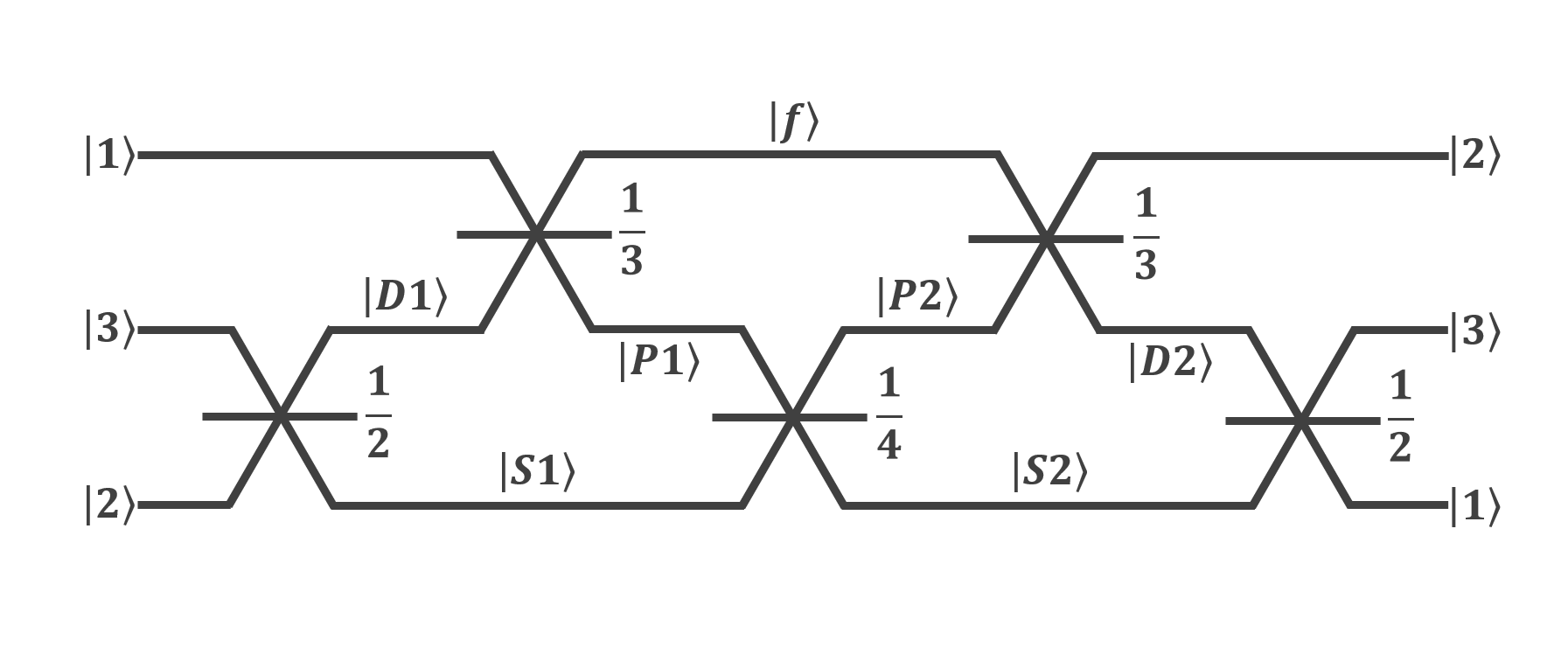}
 \caption{{\bf Schematic diagram of Hofmann's three-path interferometer for testing quantum contextuality \cite{Hofmann2023}.} $\left| i \right\rangle$ denotes the eigenstate of a photon travelling along each path. Consequently, a photon incident from input port $\left|1\right\rangle$, $\left|2\right\rangle$, or $\left|3\right\rangle$ exits from the corresponding output port $\left|1\right\rangle$, $\left|2\right\rangle$, or $\left|3\right\rangle$, respectively.
}
 \label{fig.Three Path Interferometer}
\end{figure}

Figure~\ref{fig.Three Path Interferometer} presents a schematic of the three-path interferometer designed by Hofmann to examine quantum contextuality. 
The interferometer has three input and three output ports, comprising a total of five beam splitters. 
The reflectivities of these beam splitters are individually set to $1/2$, $1/3$, and $1/4$. 
Under these conditions, a photon injected into one of the input modes on the left — specifically $|1\rangle$, $|2\rangle$, or $|3\rangle$ — is deterministically routed to the corresponding output mode on the right, namely $|1\rangle$, $|2\rangle$, or $|3\rangle$.
The photon's trajectory can be inferred from measurement outcomes associated with the three mutually orthogonal paths. 
For instance, the set of input modes $\{1,2,3\}$ constitutes a single measurement context, and interference between paths $|2\rangle$ and $|3\rangle$ at the first beam splitter gives rise to a second distinct context, $\{1, D1, S1\}$.
In the same way, the following 4 beam splitters each sequentially give rise to another distinct context ($\{f, P1, S1\}$, $\{f, P2, S2\}$, $\{2, D2, S2\}$) before returning us to $\{1,2,3\}$.
Consequently, the interferometric system illustrated in Fig.~\ref{fig.Three Path Interferometer} comprises five distinct contexts: $\{1,2,3\}$, $\{1, D1, S1\}$, $\{f, P1, S1\}$, $\{f, P2, S2\}$, and $\{2, D2, S2\}$.

\begin{figure*}[t]
  \centering
  \includegraphics[width=\textwidth]{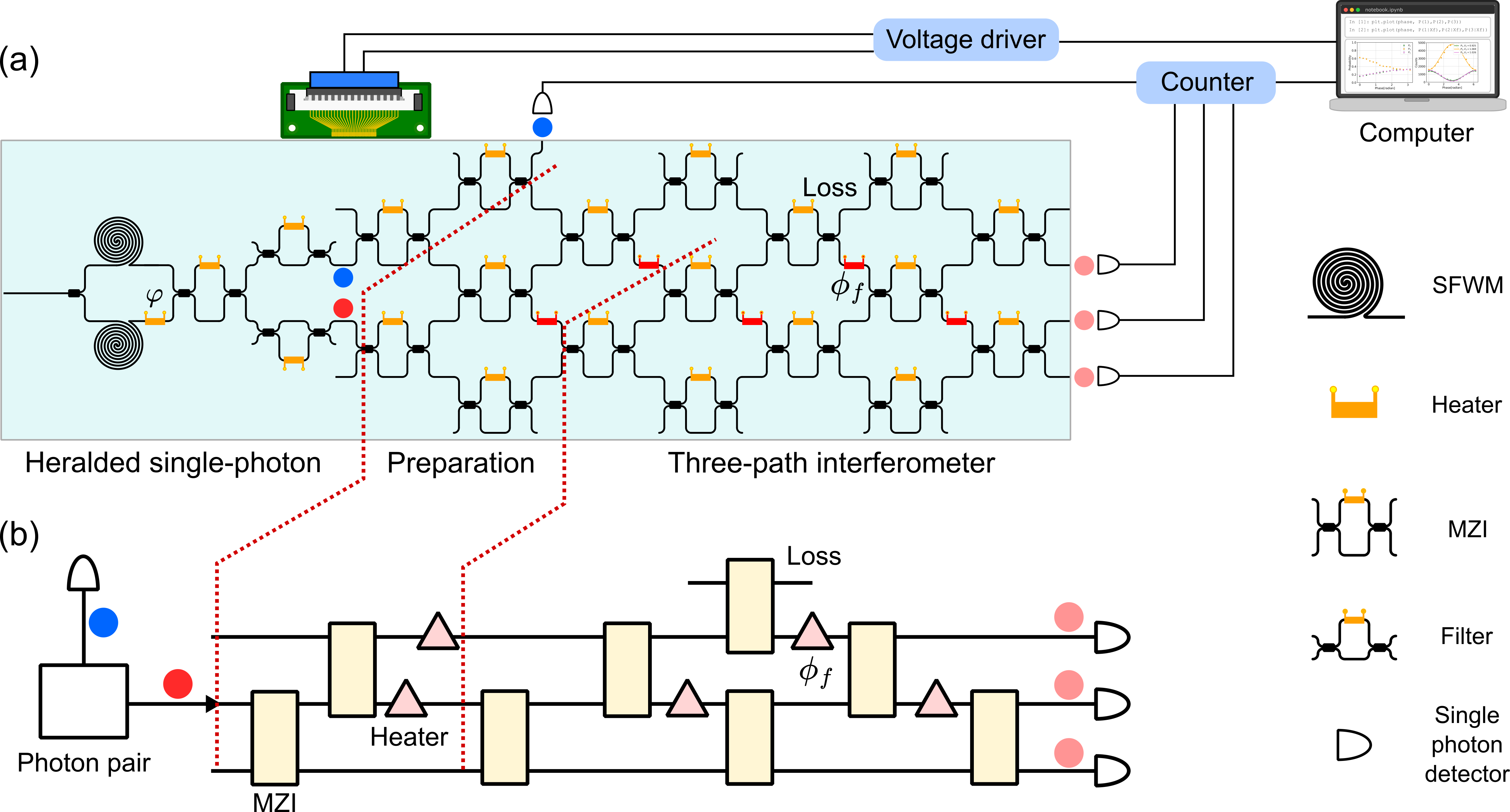}
  \caption{{\bf Experimental setup for demonstrating counterfactual gain-linked quantum contextuality in a silicon integrated circuit.} Fig.~2(a): Overview of the fabricated silicon photonic integrated circuit. The circuit consists of three parts: Heralded single-photon, Preparation, and Three-path interferometer.
  First, a telecommunications-wavelength laser is used as the pump light and injected into the chip. Through spontaneous four-wave mixing (SFWM), signal and idler photons are generated. The residual pump light is removed immediately after SFWM by a frequency filter.
One of the generated photons is used as a heralding signal, while the other is injected into the subsequent circuit as a heralded single photon. In the Preparation section, arbitrary quantum states are generated, and quantum contextuality is verified in the Three-path interferometer.
The entire chip is temperature-stabilised to operate at room temperature. The output photons are detected using superconducting nanowire single-photon detectors (SNSPDs).
Fig.~2(b): Simplified circuit diagram. The heaters shown in Fig.~2(b) correspond to the heaters indicated in red in Fig.~2(a).}
  \label{fig.ExperimentSetup}
\end{figure*}

As shown in the abstract in \cite{Cabello2013,Ji2024quantitative}, and in the three-path interferometer in \cite{Hofmann2023}, in the framework of noncontextual theories, the photon's propagation path through the interferometer (and so probabilities of detection at different points during the interferometer) is assumed to be fixed for all five measurement contexts. Specifically, when a photon traverses the path $|f\rangle$, its trajectory is uniquely constrained. For example, if a photon enters via path $|1\rangle$ and exits from the same mode after passing through $|f\rangle$, in a noncontextual model, it must necessarily have propagated through path $|D2\rangle$. Likewise, photons injected into the system via states $|2\rangle$ or $|3\rangle$ are required by such a model to pass through at least one of the intermediate paths $|D1\rangle$ or $|D2\rangle$. Accordingly, under any noncontextual model, the following inequality must be satisfied: 
\begin{equation} 
P(f) \leq P(D1) + P(D2) 
\label{eq:noncontextual_bound} 
\end{equation} 
This inequality represents a fundamental constraint imposed by noncontextuality, characterising the conditions under which photon trajectories remain consistent with predetermined path assignments. A violation of this inequality thus provides compelling evidence of quantum contextuality.

Noncontextual models are generally incompatible with the predictions of quantum mechanics, and a broad class of quantum states violate the inequality expressed in Eq.~\eqref{eq:noncontextual_bound}. As discussed in Ref.~\cite{Hofmann2023}, a very illustrative example is given by the quantum state formed of an equal superposition of all three input paths $|1\rangle$, $|2\rangle$, and $|3\rangle$, 
\begin{equation} 
|N_f \rangle = \frac{1}{\sqrt{3}} \left( |1\rangle + |2\rangle + |3\rangle \right) 
\label{eq:state1} 
\end{equation} 
In this state, quantum interference arises between paths $|2\rangle$ and $|3\rangle$, as well as between $|1\rangle$ and $|3\rangle$, leading to vanishing detection probabilities $P(D1) = P(D2) = 0$. Since the only path open to photons passing through $|f\rangle$ would take them from $\mid 1 \rangle$ to $\mid 2 \rangle$, it would seem that the probability of finding photons in $| f \rangle$ should be $P(f)=0$, in accordance with Eq.~\eqref{eq:noncontextual_bound}. However, for the state $|N_f\rangle$ defined in Eq.~\eqref{eq:state1} (named for our noncontextual expectation of finding no photon in $\mid f \rangle$ when we input this state into the interferometer), the probability of detecting the photon in path $|f\rangle$ is given by 
\begin{equation} 
P(f) = |\langle f | N_f \rangle|^2 = \frac{1}{9} > 0, 
\end{equation} 
clearly violating the inequality. This contradiction provides compelling evidence for quantum contextuality.

\section{Proving Quantum Contextuality Requires Counterfactual Gain}
One approach to directly test quantum contextuality, by witnessing the violation of the photon propagation inequality given by Eq.~\eqref{eq:noncontextual_bound}, would be to place detectors in paths $\left|f\right\rangle$, $\left|D1\right\rangle$, and $\left|D2\right\rangle$ of the system shown in Fig.~\ref{fig.Three Path Interferometer}, without modifying the setup, and measuring the corresponding probabilities. However, depending on the implementation of the system, it may be technically challenging to place detectors directly in these paths.

An alternative approach to verify the violation of the inequality in Eq.~\eqref{eq:noncontextual_bound} would be to look at the effects of modifications to a specific path on the output statistics of the interferometer. For instance, by placing an absorber in an intermediate path of the interferometer and comparing the output probability distributions with and without the absorber, we would expect to be able to infer whether the photon has passed through the path where the absorber is located. This would allow us to examine the relationship between the output and the photon that traversed the path with the absorber, thereby enabling indirect identification of the photon's propagation path.

Such inference relies on the assumption that, if the photon does not pass through path $\left|f\right\rangle$, the presence of the absorber will not affect the output probabilities. However, as shown in Ref.~\cite{Hance2024}, quantum theory predicts that the effect of an absorber goes beyond the mere loss of photons. It also induces a change in the output probability distribution, corresponding to a re-distribution of a number of photons that can be greater than the number of absorbed photons. This phenomenon, known as \textit{counterfactual gain}, results in an apparent contradiction with the assumption that only the absorbed photons interacted with the absorber. On the practical side, this means that the presence of an absorber can be deduced from the distribution of photons detected in the output ports, increasing the chance of success beyond the mere probability of absorption. 

The relation given in Ref.~\cite{Hance2024} shows that counterfactual gain is enhanced by negative quasiprobabilities, similar to the ones responsible for the violation of the noncontextual inequality in \cite{Hofmann2024}. Based on this relation, we investigated the counterfactual gain required for a violation of the inequality given by Eq.~\eqref{eq:noncontextual_bound}. The \textit{counterfactual gain} is defined as the total amount by which the output probability increases when comparing the cases with and without an absorber placed along the particle’s propagation path. Specifically, in the system shown in Fig.~1, when an absorber is placed in path $\left|f\right\rangle$ within the interferometer, the difference between the left- and right-hand sides of Eq.~\eqref{eq:noncontextual_bound} - serving as an indicator of contextuality — can be expressed in terms of the output probability distributions as follows:
\begin{equation}
\begin{aligned}
P(f)& - P(D1) - P(D2) = 
\\&\left[ P(3|X_f) - P(3) \right]
- \frac{1}{2} \left[ P(1|X_f) + P(2|X_f) \right]
\end{aligned}
\label{eq:contextual_difference}
\end{equation}
Here, $P(i|X_f)$ denotes the probability of detecting a photon at output port $\left|i\right\rangle$ ($i = 1, 2, 3$) when an absorber is placed in path $\left|f\right\rangle$. If the value given by Eq.~\eqref{eq:contextual_difference} is positive, the inequality is violated and quantum contextuality has been confirmed.

\begin{figure*}[t]
  \centering
  \includegraphics[width=\textwidth]{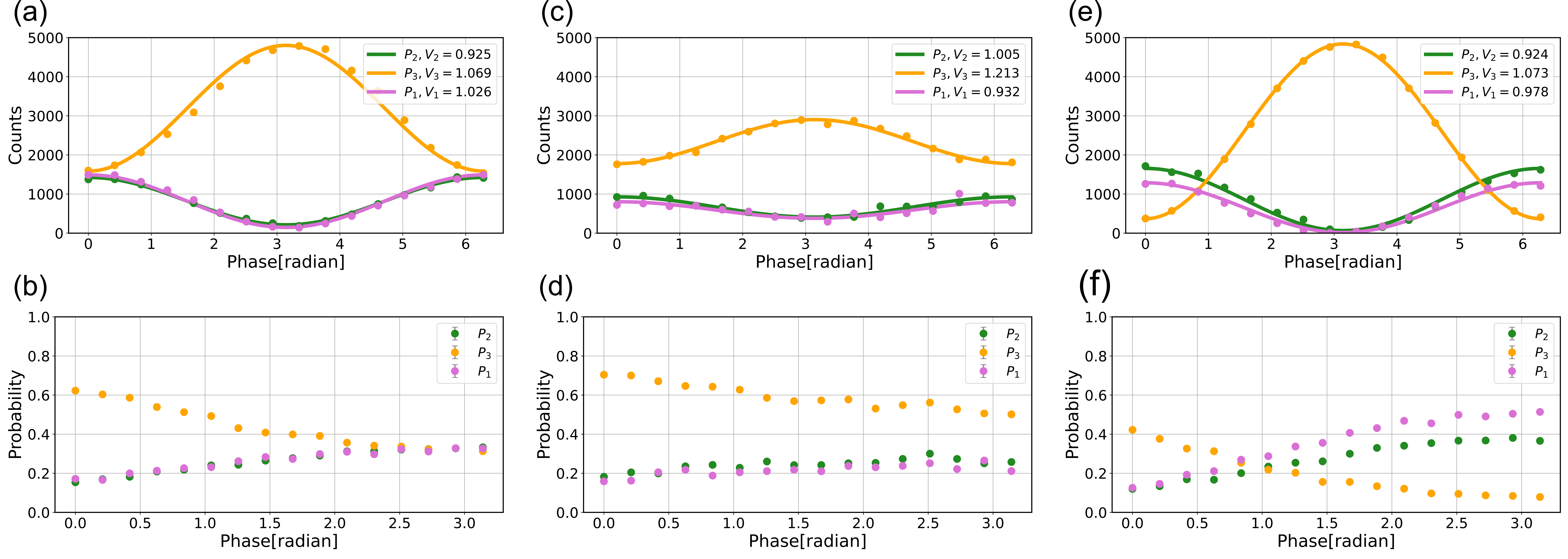}
  \caption{{\bf Experimental results demonstrating quantum contextuality.} Figures~3(a) and 3(b) present the results obtained using the input state $| N_f \rangle$, Figs.~3(c) and 3(d) correspond to the input state $| B_f \rangle$, and Figs.~3(e) and 3(f) correspond to the input state $| V_0 \rangle$. Figures~3(a), 3(c), and 3(e) show the variations in the output probability distributions when the phase of the phase shifter placed in path $\left|f\right\rangle$ was varied. Figures~3(b), 3(d), and 3(f) display the changes in the output probability distributions when the transmittance of path $\left|f\right\rangle$ was varied. The horizontal axis in Figs.~3(b), 3(d), and 3(f) represents the phase of the Mach-Zehnder interferometer (MZI) placed in path $\left|f\right\rangle$. When the MZI phase is 0, the transmittance of path $\left|f\right\rangle$ is zero; when the phase is $\pi$, the transmittance reaches unity. Each data point was obtained by integrating the measurement over a period of 100 seconds. The error bars were calculated under the assumption that each data point follows a Poisson distribution.
}
  \label{fig.ExperimentResults}
\end{figure*}

In the system shown in Fig.~\ref{fig.Three Path Interferometer}, the first term in Eq.~\eqref{eq:contextual_difference}, $P(3|X_f) - P(3)$, corresponds to the counterfactual gain in output $| 3 \rangle$ when path $| f \rangle$ is blocked. Since the second term, $\frac{1}{2} \left( P(1|X_f) + P(2|X_f) \right)$, is a non-negative sum of probabilities accounting for additional counts in outputs $| 1 \rangle$ and $| 2 \rangle$,  Eq.~\eqref{eq:contextual_difference} can only yield a positive value if the counterfactual gain is also positive. We can therefore conclude that counterfactual gain is necessary for the observation of quantum contextuality. At the same time, counterfactual gain cannot guarantee the observation of quantum contextuality, since it could be smaller than the average of the additional output probabilities $\frac{1}{2} \left( P(1|X_f) + P(2|X_f) \right)$. Counterfactual gain is necessary but not sufficient for quantum contextuality. An observation of quantum contextuality always guarantees a minimal amount of counterfactual gain, so that the violation of the inequality in Eq.~\eqref{eq:noncontextual_bound} serves as an indicator that the quantum state is highly sensitive to the presence of an absorber in path $\mid f \rangle$.

\section{Demonstrating Quantum Contextuality via Counterfactual Gain Using a Silicon Photonic Chip}
We now experimentally demonstrate quantum contextuality using a silicon photonic integrated circuit. Our chip was fabricated via a commercially available multi-project wafer (MPW) service (Fig.~\ref{fig.ExperimentSetup}) and integrates a photon-pair source based on spontaneous four-wave mixing (SFWM) with a universal $4 \times 4$ unitary transformation circuit on a silicon platform.

In the experiment, three of the four available spatial modes were utilised for verification. A heralded single-photon source was employed: photon pairs were generated through the third-order nonlinear optical response of silicon, with one photon serving as a trigger signal and the other being injected into the unitary circuit as a heralded single photon. This configuration guarantees that each detection event corresponds to the injection of exactly one photon into the transformation circuit.

An arbitrary three-dimensional quantum state was prepared using two Mach-Zehnder interferometers (MZIs) and two phase shifters, arranged prior to the 3-path interferometer as indicated in the “Preparation” section of Fig.~\ref{fig.ExperimentSetup}. The five beam splitters with distinct reflectivities that form the 3-path interferometer were implemented using MZIs configured as tunable beam splitters. Phase shifters placed after each MZI were used to compensate for additional phases introduced by the interferometers.

We prepared several distinct input states and examined quantum contextuality for each. Figs.~\ref{fig.ExperimentResults}(a) and \ref{fig.ExperimentResults}(b) present the results obtained for the input state $\left| N_f \right\rangle = \left( \left|1\right\rangle + \left|2\right\rangle + \left|3\right\rangle \right)/\sqrt{3}$. Fig.~\ref{fig.ExperimentResults}(a) shows the variation in the output probability distribution as the phase $\phi_f$ applied to the phase shifter in path $\left|f\right\rangle$ was modulated, while Fig.~\ref{fig.ExperimentResults}(b) presents the output distributions when the transmittance of the same path was gradually tuned. The transmittance was controlled via a tunable beam splitter, realized using a Mach-Zehnder interferometer (MZI) combined with a phase shifter to compensate for the additional phase induced by the MZI.

According to noncontextual theories, the probability that a photon traverses path $\left|f\right\rangle$ for the state $\left| N_f \right\rangle$ is zero. Thus, operations on this path should not affect the output.
However, as shown in Fig.~\ref{fig.ExperimentResults}(a), the output distribution exhibits significant sensitivity to variations in $\phi_f$, clearly indicating the presence of a photon in path $\left|f\right\rangle$.
From the observed interference fringes, the output probabilities at $\phi_f = 0$ were $P(1) = 0.347$, $P(2) = 0.331$, and $P(3) = 0.322$, normalized to ensure their sum is unity.

Similarly, Fig.~\ref{fig.ExperimentResults}(b) shows that decreasing the transmittance of path $\left|f\right\rangle$ led to a notable increase in $P(3)$ and corresponding decreases in $P(1)$ and $P(2)$.
When the path was fully blocked, the measured output probabilities were $P(1|X_f) = 0.161$, $P(2|X_f) = 0.144$, and $P(3|X_f) = 0.585$.
The counterfactual gain at output port 3 was thus $P(3|X_f) - P(3) = 0.263$, providing strong evidence of nonclassical behaviour.

The degree of violation of the inequality defined in Eq.~\eqref{eq:contextual_difference} was calculated as: 
\begin{equation} \left(P(3|X_f) - P(3)\right) - \frac{1}{2} \left(P(1|X_f) + P(2|X_f)\right) = 0.11,
\end{equation}
corresponding to an 11\% violation and thus confirming quantum contextuality.

However, the presence of counterfactual gain does not necessarily imply contextuality.
Figs.~\ref{fig.ExperimentResults}(c) and \ref{fig.ExperimentResults}(d) show results obtained with a state close to the boundary between violation and non-violation of the inequality. This state is given by
\begin{equation}
    \left| B_f \right\rangle = \frac{1}{\sqrt{17}}\left( 2\left|1\right\rangle + 2\left|2\right\rangle + 3\left|3\right\rangle \right),
\end{equation}
where the phase and transmittance of path $\left|f\right\rangle$ were varied.
In Fig.~\ref{fig.ExperimentResults}(d), an increase in $P(3)$ was observed upon inserting the absorber.
Without the absorber, the measured probabilities were $P(1) = 0.238$, $P(2) = 0.243$, and $P(3) = 0.518$; with the absorber, they were $P(1|X_f) = 0.144$, $P(2|X_f) = 0.165$, and $P(3|X_f) = 0.638$.
Thus, the counterfactual gain was $P(3|X_f) - P(3) = 0.120$.
Nevertheless, the inequality evaluation yields: 
\begin{equation}
(P(3|X_f) - P(3)) - \frac{1}{2}(P(1|X_f) + P(2|X_f)) = -0.04,
\end{equation}
showing no violation and hence no evidence of quantum contextuality.
The results for this input state demonstrate that counterfactual gain alone does not constitute proof of contextual behaviour.

Finally, Figs.~\ref{fig.ExperimentResults}(e) and \ref{fig.ExperimentResults}(f) present the results for a state that is close to the maximal violation of the inequality, 
\begin{equation}
    \left| V_0 \right\rangle = \frac{1}{3}\left( 2\left|1\right\rangle + 2\left|2\right\rangle + \left|3\right\rangle \right).
\end{equation}
Without the absorber, the output probabilities were $P(1) = 0.465$, $P(2) = 0.452$, and $P(3) = 0.083$; with the absorber, they changed to $P(1|X_f) = 0.125$, $P(2|X_f) = 0.119$, and $P(3|X_f) = 0.419$.
The counterfactual gain at output port 3 was $P(3|X_f) - P(3) = 0.336$, and the inequality was evaluated as:
\begin{equation}
(P(3|X_f) - P(3)) - \frac{1}{2}(P(1|X_f) + P(2|X_f)) = 0.214,
\end{equation}
corresponding to a violation exceeding 21\%, thus demonstrating that compelling experimental evidence for quantum contextuality is indicative of extremely high counterfactual gain.

\section{Discussion}
We have shown that quantum contextuality requires counterfactual gain, indicating that input states that maximally violate noncontextual inequalities also provide maximal resources for the detection of absorbers using counterfactual gain. Our results show that quantum contextuality can be characterized entirely in terms of counterfactual gain, linking the paradoxical aspects of contextual statistics directly with the quantum advantage of counterfactual gain. Through the precise implementation of a three-path interferometer on a silicon photonic integrated circuit, we demonstrated that the insertion of absorbers in selected paths leads to output probability changes that reflect the contextual nature of quantum mechanics. Our findings establish counterfactual gain as a practical and accessible indicator of quantum contextuality, providing a new route for its direct experimental verification. The integrated photonic platform employed herein offers enhanced stability and scalability, suggesting promising opportunities for exploring contextuality as a fundamental resource in quantum information science. Future directions include extending the framework to higher-dimensional systems and applying it to tasks such as quantum metrology, quantum computation and quantum communication.

\section{Methods}
\subsection{Heralded single photon source}
In this study, due to technical constraints, we employed a probabilistic method to generate heralded single photons. A continuous-wave pump laser with a wavelength of 1549.32nm was split into two paths using a beam splitter. Each beam was injected into a silicon waveguide approximately 1.4cm in length, where signal and idler photon pairs were generated via spontaneous four-wave mixing (SFWM). The generated photons had central wavelengths of approximately $1559.0 \pm 0.8$~nm (signal) and $1539.8 \pm 0.8$~nm (idler), as determined by spectral filtering.

To control the interference between the two photon-pair generation paths, a relative phase shift $\varphi$ was applied to one of the arms, and the two paths were subsequently recombined at a second beam splitter. As shown in previous work~\cite{Silverstone2014}, under the condition of low pump power—where the probability of multi-pair generation is negligible—the output state can be approximated as a coherent superposition of two components: a bunching state $|\Psi_B\rangle$, in which both photons exit in the same spatial mode, and an antibunching state $|\Psi_A\rangle$, in which the photons are separated into different spatial modes:
\begin{equation} 
|\psi\rangle \approx \cos \varphi |\Psi_B\rangle + \sin \varphi |\Psi_A\rangle 
\end{equation}

In the experiment, the phase was set to $\varphi = \pi/2$ to maximize the antibunching contribution, thereby ensuring that the signal and idler photons exit into distinct spatial modes. The antibunching state is given by:
\begin{equation} |\Psi_A\rangle = \frac{1}{\sqrt{2}} \left( |i; s\rangle_{ab} + |s; i\rangle_{ab} \right), \end{equation}
where $i$ and $s$ denote "idler" and "signal", and $a$ and $b$ represent the two spatial output modes. In the experiment, one photon (typically the signal) was detected and used as a heralding trigger, while the other (idler) photon was directed into the subsequent circuit as a heralded single photon, with a heralding probability of approximately 50\%.

\subsection{Interference visibility}
Figure~\ref{fig.ExperimentResults} shows the changes in the output probability distributions when the phase $\phi_f$ applied to the phase shifter in path $f$ is varied. Each dataset was fitted using theoretical models assuming the input states to be $|N_f\rangle$, $|B_f\rangle$, or $|V_0\rangle$, incorporating the interference visibility $V$ into the phase-dependent probability expressions.

For the state $|N_f\rangle$, the output probabilities as functions of $\phi_f$ are given by:
\begin{eqnarray} 
P(1|\phi_f) &=& \frac{5}{27} + \frac{4}{27} V_1 \cos \phi_f \nonumber\\
P(2|\phi_f) &=& \frac{5}{27} + \frac{4}{27} V_2 \cos \phi_f \nonumber\\
P(3|\phi_f) &=& \frac{17}{27} - \frac{8}{27} V_3 \cos \phi_f 
\end{eqnarray}
The experimentally obtained visibilities were $V_1 = 1.03$, $V_2 = 0.93$, and $V_3 = 1.07$.

For the state $|B_f\rangle$, the phase dependence of the output probabilities is:
\begin{eqnarray} 
P(1|\phi_f) &=& \frac{26}{153} + \frac{10}{153} V_1 \cos \phi_f \nonumber\\
P(2|\phi_f) &=& \frac{26}{153} + \frac{10}{153} V_2 \cos \phi_f \nonumber\\
P(3|\phi_f) &=& \frac{101}{153} - \frac{20}{153} V_3 \cos \phi_f 
\end{eqnarray}
In this case, the visibilities were $V_1 = 0.93$, $V_2 = 1.01$, and $V_3 = 1.21$.

For the state $|V_0\rangle$, the output probabilities follow:
\begin{eqnarray} 
P(1|\phi_f) &=& \frac{2}{9} + \frac{2}{9} V_1 \cos \phi_f \nonumber\\
P(2|\phi_f) &=& \frac{2}{9} + \frac{2}{9} V_2 \cos \phi_f \nonumber\\
P(3|\phi_f) &=& \frac{5}{9} - \frac{4}{9} V_3 \cos \phi_f 
\end{eqnarray}
The corresponding visibilities were $V_1 = 0.98$, $V_2 = 0.92$, and $V_3 = 1.07$.

\textbf{\emph{Acknowledgements}}

\textbf{Funding:}
This work was supported by JST PRESTO Grant No. JPMJPR1864, and ERATO, Japan Science and Technology Agency (JPMJER2402), JSPS KAKENHI Grant Number JP24K00559. JRH acknowledges support from a Royal Society Research Grant (RG/R1/251590).\\


\begin{thebibliography}{25}%
\makeatletter
\providecommand \@ifxundefined [1]{%
 \@ifx{#1\undefined}
}%
\providecommand \@ifnum [1]{%
 \ifnum #1\expandafter \@firstoftwo
 \else \expandafter \@secondoftwo
 \fi
}%
\providecommand \@ifx [1]{%
 \ifx #1\expandafter \@firstoftwo
 \else \expandafter \@secondoftwo
 \fi
}%
\providecommand \natexlab [1]{#1}%
\providecommand \enquote  [1]{``#1''}%
\providecommand \bibnamefont  [1]{#1}%
\providecommand \bibfnamefont [1]{#1}%
\providecommand \citenamefont [1]{#1}%
\providecommand \href@noop [0]{\@secondoftwo}%
\providecommand \href [0]{\begingroup \@sanitize@url \@href}%
\providecommand \@href[1]{\@@startlink{#1}\@@href}%
\providecommand \@@href[1]{\endgroup#1\@@endlink}%
\providecommand \@sanitize@url [0]{\catcode `\\12\catcode `\$12\catcode `\&12\catcode `\#12\catcode `\^12\catcode `\_12\catcode `\%12\relax}%
\providecommand \@@startlink[1]{}%
\providecommand \@@endlink[0]{}%
\providecommand \url  [0]{\begingroup\@sanitize@url \@url }%
\providecommand \@url [1]{\endgroup\@href {#1}{\urlprefix }}%
\providecommand \urlprefix  [0]{URL }%
\providecommand \Eprint [0]{\href }%
\providecommand \doibase [0]{https://doi.org/}%
\providecommand \selectlanguage [0]{\@gobble}%
\providecommand \bibinfo  [0]{\@secondoftwo}%
\providecommand \bibfield  [0]{\@secondoftwo}%
\providecommand \translation [1]{[#1]}%
\providecommand \BibitemOpen [0]{}%
\providecommand \bibitemStop [0]{}%
\providecommand \bibitemNoStop [0]{.\EOS\space}%
\providecommand \EOS [0]{\spacefactor3000\relax}%
\providecommand \BibitemShut  [1]{\csname bibitem#1\endcsname}%
\let\auto@bib@innerbib\@empty
\bibitem [{\citenamefont {Kochen}\ and\ \citenamefont {Specker}(1967)}]{KochenSpecker1967}%
  \BibitemOpen
  \bibfield  {author} {\bibinfo {author} {\bibfnamefont {S.}~\bibnamefont {Kochen}}\ and\ \bibinfo {author} {\bibfnamefont {E.}~\bibnamefont {Specker}},\ }\href@noop {} {\bibfield  {journal} {\bibinfo  {journal} {Indiana Univ. Math. J.}\ }\textbf {\bibinfo {volume} {17}},\ \bibinfo {pages} {59} (\bibinfo {year} {1967})}\BibitemShut {NoStop}%
\bibitem [{\citenamefont {Budroni}\ \emph {et~al.}(2022)\citenamefont {Budroni}, \citenamefont {Cabello}, \citenamefont {G{\"{u}}hne}, \citenamefont {Kleinmann},\ and\ \citenamefont {Larsson}}]{Budroni2022}%
  \BibitemOpen
  \bibfield  {author} {\bibinfo {author} {\bibfnamefont {C.}~\bibnamefont {Budroni}}, \bibinfo {author} {\bibfnamefont {A.}~\bibnamefont {Cabello}}, \bibinfo {author} {\bibfnamefont {O.}~\bibnamefont {G{\"{u}}hne}}, \bibinfo {author} {\bibfnamefont {M.}~\bibnamefont {Kleinmann}},\ and\ \bibinfo {author} {\bibfnamefont {J.~{\AA}.}\ \bibnamefont {Larsson}},\ }\href {https://doi.org/10.1103/RevModPhys.94.045007} {\bibfield  {journal} {\bibinfo  {journal} {Reviews of Modern Physics}\ }\textbf {\bibinfo {volume} {94}},\ \bibinfo {pages} {1} (\bibinfo {year} {2022})}\BibitemShut {NoStop}%
\bibitem [{\citenamefont {Hardy}(1992)}]{Hardy1992}%
  \BibitemOpen
  \bibfield  {author} {\bibinfo {author} {\bibfnamefont {L.}~\bibnamefont {Hardy}},\ }\href@noop {} {\bibfield  {journal} {\bibinfo  {journal} {Phys. Rev. Lett.}\ }\textbf {\bibinfo {volume} {68}},\ \bibinfo {pages} {2981} (\bibinfo {year} {1992})}\BibitemShut {NoStop}%
\bibitem [{\citenamefont {Hardy}(1993)}]{Hardy1993}%
  \BibitemOpen
  \bibfield  {author} {\bibinfo {author} {\bibfnamefont {L.}~\bibnamefont {Hardy}},\ }\href@noop {} {\bibfield  {journal} {\bibinfo  {journal} {Phys. Rev. Lett.}\ }\textbf {\bibinfo {volume} {71}},\ \bibinfo {pages} {1665} (\bibinfo {year} {1993})}\BibitemShut {NoStop}%
\bibitem [{\citenamefont {Klyachko}\ \emph {et~al.}(2008)\citenamefont {Klyachko}, \citenamefont {Can}, \citenamefont {Binicio^^c4^^9flu},\ and\ \citenamefont {Shumovsky}}]{Klyachko2008}%
  \BibitemOpen
  \bibfield  {author} {\bibinfo {author} {\bibfnamefont {A.~A.}\ \bibnamefont {Klyachko}}, \bibinfo {author} {\bibfnamefont {M.~A.}\ \bibnamefont {Can}}, \bibinfo {author} {\bibfnamefont {S.}~\bibnamefont {Binicio^^c4^^9flu}},\ and\ \bibinfo {author} {\bibfnamefont {A.~S.}\ \bibnamefont {Shumovsky}},\ }\href@noop {} {\bibfield  {journal} {\bibinfo  {journal} {Phys. Rev. Lett.}\ }\textbf {\bibinfo {volume} {101}},\ \bibinfo {pages} {020403} (\bibinfo {year} {2008})}\BibitemShut {NoStop}%
\bibitem [{\citenamefont {Cabello}\ \emph {et~al.}(2013)\citenamefont {Cabello}, \citenamefont {Badzi\c{a}g}, \citenamefont {{Terra Cunha}},\ and\ \citenamefont {Bourennane}}]{Cabello2013}%
  \BibitemOpen
  \bibfield  {author} {\bibinfo {author} {\bibfnamefont {A.}~\bibnamefont {Cabello}}, \bibinfo {author} {\bibfnamefont {P.}~\bibnamefont {Badzi\c{a}g}}, \bibinfo {author} {\bibfnamefont {M.}~\bibnamefont {{Terra Cunha}}},\ and\ \bibinfo {author} {\bibfnamefont {M.}~\bibnamefont {Bourennane}},\ }\href {https://doi.org/10.1103/PhysRevLett.111.180404} {\bibfield  {journal} {\bibinfo  {journal} {Physical Review Letters}\ }\textbf {\bibinfo {volume} {111}},\ \bibinfo {pages} {3} (\bibinfo {year} {2013})}\BibitemShut {NoStop}%
\bibitem [{\citenamefont {Spekkens}(2005)}]{Spekkens2005}%
  \BibitemOpen
  \bibfield  {author} {\bibinfo {author} {\bibfnamefont {R.~W.}\ \bibnamefont {Spekkens}},\ }\bibfield  {journal} {\bibinfo  {journal} {Physical Review A - Atomic, Molecular, and Optical Physics}\ }\textbf {\bibinfo {volume} {71}},\ \href {https://doi.org/10.1103/PhysRevA.71.052108} {10.1103/PhysRevA.71.052108} (\bibinfo {year} {2005})\BibitemShut {NoStop}%
\bibitem [{\citenamefont {Tezzin}\ \emph {et~al.}(2025)\citenamefont {Tezzin}, \citenamefont {Amaral},\ and\ \citenamefont {Hance}}]{tezzin2025ontologicalmodelsadequatelyrepresent}%
  \BibitemOpen
  \bibfield  {author} {\bibinfo {author} {\bibfnamefont {A.}~\bibnamefont {Tezzin}}, \bibinfo {author} {\bibfnamefont {B.}~\bibnamefont {Amaral}},\ and\ \bibinfo {author} {\bibfnamefont {J.~R.}\ \bibnamefont {Hance}},\ }\href {https://arxiv.org/abs/2502.15615} {\bibinfo {title} {Ontological models cannot adequately represent state update for sequential measurement of incompatible observables}} (\bibinfo {year} {2025}),\ \Eprint {https://arxiv.org/abs/2502.15615} {arXiv:2502.15615 [quant-ph]} \BibitemShut {NoStop}%
\bibitem [{\citenamefont {Huang}\ \emph {et~al.}(2003)\citenamefont {Huang}, \citenamefont {Li}, \citenamefont {Zhang}, \citenamefont {Pan},\ and\ \citenamefont {Guo}}]{Huang2003}%
  \BibitemOpen
  \bibfield  {author} {\bibinfo {author} {\bibfnamefont {Y.-f.}\ \bibnamefont {Huang}}, \bibinfo {author} {\bibfnamefont {C.-f.}\ \bibnamefont {Li}}, \bibinfo {author} {\bibfnamefont {Y.-s.}\ \bibnamefont {Zhang}}, \bibinfo {author} {\bibfnamefont {J.-w.}\ \bibnamefont {Pan}},\ and\ \bibinfo {author} {\bibfnamefont {G.-c.}\ \bibnamefont {Guo}},\ }\href@noop {} {\bibfield  {journal} {\bibinfo  {journal} {Physical Review Letters}\ }\textbf {\bibinfo {volume} {90}},\ \bibinfo {pages} {250401} (\bibinfo {year} {2003})}\BibitemShut {NoStop}%
\bibitem [{\citenamefont {D'Ambrosio}\ \emph {et~al.}(2013)\citenamefont {D'Ambrosio}, \citenamefont {Herbauts}, \citenamefont {Amselem}, \citenamefont {Nagali}, \citenamefont {Bourennane}, \citenamefont {Sciarrino},\ and\ \citenamefont {Cabello}}]{DAmbrosio2013}%
  \BibitemOpen
  \bibfield  {author} {\bibinfo {author} {\bibfnamefont {V.}~\bibnamefont {D'Ambrosio}}, \bibinfo {author} {\bibfnamefont {I.}~\bibnamefont {Herbauts}}, \bibinfo {author} {\bibfnamefont {E.}~\bibnamefont {Amselem}}, \bibinfo {author} {\bibfnamefont {E.}~\bibnamefont {Nagali}}, \bibinfo {author} {\bibfnamefont {M.}~\bibnamefont {Bourennane}}, \bibinfo {author} {\bibfnamefont {F.}~\bibnamefont {Sciarrino}},\ and\ \bibinfo {author} {\bibfnamefont {A.}~\bibnamefont {Cabello}},\ }\href {https://doi.org/10.1103/PhysRevX.3.011012} {\bibfield  {journal} {\bibinfo  {journal} {Physical Review X}\ }\textbf {\bibinfo {volume} {3}},\ \bibinfo {pages} {1} (\bibinfo {year} {2013})}\BibitemShut {NoStop}%
\bibitem [{\citenamefont {Zhang}\ \emph {et~al.}(2019)\citenamefont {Zhang}, \citenamefont {Xu}, \citenamefont {Xie}, \citenamefont {Zhang}, \citenamefont {Smith}, \citenamefont {Kim},\ and\ \citenamefont {Zhang}}]{Zhang2019}%
  \BibitemOpen
  \bibfield  {author} {\bibinfo {author} {\bibfnamefont {A.}~\bibnamefont {Zhang}}, \bibinfo {author} {\bibfnamefont {H.}~\bibnamefont {Xu}}, \bibinfo {author} {\bibfnamefont {J.}~\bibnamefont {Xie}}, \bibinfo {author} {\bibfnamefont {H.}~\bibnamefont {Zhang}}, \bibinfo {author} {\bibfnamefont {B.~J.}\ \bibnamefont {Smith}}, \bibinfo {author} {\bibfnamefont {M.~S.}\ \bibnamefont {Kim}},\ and\ \bibinfo {author} {\bibfnamefont {L.}~\bibnamefont {Zhang}},\ }\href {https://doi.org/10.1103/PhysRevLett.122.080401} {\bibfield  {journal} {\bibinfo  {journal} {Physical Review Letters}\ }\textbf {\bibinfo {volume} {122}},\ \bibinfo {pages} {1} (\bibinfo {year} {2019})}\BibitemShut {NoStop}%
\bibitem [{\citenamefont {Qu}\ \emph {et~al.}(2021)\citenamefont {Qu}, \citenamefont {Wang}, \citenamefont {Xiao}, \citenamefont {Zhan},\ and\ \citenamefont {Xue}}]{Qu2021}%
  \BibitemOpen
  \bibfield  {author} {\bibinfo {author} {\bibfnamefont {D.}~\bibnamefont {Qu}}, \bibinfo {author} {\bibfnamefont {K.}~\bibnamefont {Wang}}, \bibinfo {author} {\bibfnamefont {L.}~\bibnamefont {Xiao}}, \bibinfo {author} {\bibfnamefont {X.}~\bibnamefont {Zhan}},\ and\ \bibinfo {author} {\bibfnamefont {P.}~\bibnamefont {Xue}},\ }\href {https://doi.org/10.1038/s41534-021-00492-1} {\bibfield  {journal} {\bibinfo  {journal} {npj Quantum Information}\ }\textbf {\bibinfo {volume} {7}},\ \bibinfo {pages} {1} (\bibinfo {year} {2021})}\BibitemShut {NoStop}%
\bibitem [{\citenamefont {Liu}\ \emph {et~al.}(2023)\citenamefont {Liu}, \citenamefont {Meng}, \citenamefont {Xu}, \citenamefont {Zhou}, \citenamefont {Chen}, \citenamefont {Xu}, \citenamefont {Li}, \citenamefont {Guo},\ and\ \citenamefont {Cabello}}]{Liu2023}%
  \BibitemOpen
  \bibfield  {author} {\bibinfo {author} {\bibfnamefont {Z.~H.}\ \bibnamefont {Liu}}, \bibinfo {author} {\bibfnamefont {H.~X.}\ \bibnamefont {Meng}}, \bibinfo {author} {\bibfnamefont {Z.~P.}\ \bibnamefont {Xu}}, \bibinfo {author} {\bibfnamefont {J.}~\bibnamefont {Zhou}}, \bibinfo {author} {\bibfnamefont {J.~L.}\ \bibnamefont {Chen}}, \bibinfo {author} {\bibfnamefont {J.~S.}\ \bibnamefont {Xu}}, \bibinfo {author} {\bibfnamefont {C.~F.}\ \bibnamefont {Li}}, \bibinfo {author} {\bibfnamefont {G.~C.}\ \bibnamefont {Guo}},\ and\ \bibinfo {author} {\bibfnamefont {A.}~\bibnamefont {Cabello}},\ }\bibfield  {journal} {\bibinfo  {journal} {Physical Review Letters}\ }\textbf {\bibinfo {volume} {130}},\ \href {https://doi.org/10.1103/PhysRevLett.130.240202} {10.1103/PhysRevLett.130.240202} (\bibinfo {year} {2023})\BibitemShut {NoStop}%
\bibitem [{\citenamefont {Kirchmair}\ \emph {et~al.}(2009)\citenamefont {Kirchmair}, \citenamefont {Z{\"{a}}hringer}, \citenamefont {Gerritsma}, \citenamefont {Kleinmann}, \citenamefont {G{\"{u}}hne}, \citenamefont {Cabello}, \citenamefont {Blatt},\ and\ \citenamefont {Roos}}]{Kirchmair2009}%
  \BibitemOpen
  \bibfield  {author} {\bibinfo {author} {\bibfnamefont {G.}~\bibnamefont {Kirchmair}}, \bibinfo {author} {\bibfnamefont {F.}~\bibnamefont {Z{\"{a}}hringer}}, \bibinfo {author} {\bibfnamefont {R.}~\bibnamefont {Gerritsma}}, \bibinfo {author} {\bibfnamefont {M.}~\bibnamefont {Kleinmann}}, \bibinfo {author} {\bibfnamefont {O.}~\bibnamefont {G{\"{u}}hne}}, \bibinfo {author} {\bibfnamefont {A.}~\bibnamefont {Cabello}}, \bibinfo {author} {\bibfnamefont {R.}~\bibnamefont {Blatt}},\ and\ \bibinfo {author} {\bibfnamefont {C.~F.}\ \bibnamefont {Roos}},\ }\href {https://doi.org/10.1038/nature08172} {\bibfield  {journal} {\bibinfo  {journal} {Nature}\ }\textbf {\bibinfo {volume} {460}},\ \bibinfo {pages} {494} (\bibinfo {year} {2009})}\BibitemShut {NoStop}%
\bibitem [{\citenamefont {Zhang}\ \emph {et~al.}(2013)\citenamefont {Zhang}, \citenamefont {Um}, \citenamefont {Zhang}, \citenamefont {An}, \citenamefont {Wang}, \citenamefont {Deng}, \citenamefont {Shen}, \citenamefont {Duan},\ and\ \citenamefont {Kim}}]{Zhang2013}%
  \BibitemOpen
  \bibfield  {author} {\bibinfo {author} {\bibfnamefont {X.}~\bibnamefont {Zhang}}, \bibinfo {author} {\bibfnamefont {M.}~\bibnamefont {Um}}, \bibinfo {author} {\bibfnamefont {J.}~\bibnamefont {Zhang}}, \bibinfo {author} {\bibfnamefont {S.}~\bibnamefont {An}}, \bibinfo {author} {\bibfnamefont {Y.}~\bibnamefont {Wang}}, \bibinfo {author} {\bibfnamefont {D.~L.}\ \bibnamefont {Deng}}, \bibinfo {author} {\bibfnamefont {C.}~\bibnamefont {Shen}}, \bibinfo {author} {\bibfnamefont {L.~M.}\ \bibnamefont {Duan}},\ and\ \bibinfo {author} {\bibfnamefont {K.}~\bibnamefont {Kim}},\ }\href {https://doi.org/10.1103/PhysRevLett.110.070401} {\bibfield  {journal} {\bibinfo  {journal} {Physical Review Letters}\ }\textbf {\bibinfo {volume} {110}},\ \bibinfo {pages} {2} (\bibinfo {year} {2013})}\BibitemShut {NoStop}%
\bibitem [{\citenamefont {Leupold}\ \emph {et~al.}(2018)\citenamefont {Leupold}, \citenamefont {Malinowski}, \citenamefont {Zhang}, \citenamefont {Negnevitsky}, \citenamefont {Alonso}, \citenamefont {Home},\ and\ \citenamefont {Cabello}}]{Leupold2018}%
  \BibitemOpen
  \bibfield  {author} {\bibinfo {author} {\bibfnamefont {F.~M.}\ \bibnamefont {Leupold}}, \bibinfo {author} {\bibfnamefont {M.}~\bibnamefont {Malinowski}}, \bibinfo {author} {\bibfnamefont {C.}~\bibnamefont {Zhang}}, \bibinfo {author} {\bibfnamefont {V.}~\bibnamefont {Negnevitsky}}, \bibinfo {author} {\bibfnamefont {J.}~\bibnamefont {Alonso}}, \bibinfo {author} {\bibfnamefont {J.~P.}\ \bibnamefont {Home}},\ and\ \bibinfo {author} {\bibfnamefont {A.}~\bibnamefont {Cabello}},\ }\href {https://doi.org/10.1103/PhysRevLett.120.180401} {\bibfield  {journal} {\bibinfo  {journal} {Physical Review Letters}\ }\textbf {\bibinfo {volume} {120}},\ \bibinfo {pages} {1} (\bibinfo {year} {2018})}\BibitemShut {NoStop}%
\bibitem [{\citenamefont {Wang}\ \emph {et~al.}(2022)\citenamefont {Wang}, \citenamefont {Zhang}, \citenamefont {Luan}, \citenamefont {Um}, \citenamefont {Wang}, \citenamefont {Qiao}, \citenamefont {Xie}, \citenamefont {Zhang}, \citenamefont {Cabello},\ and\ \citenamefont {Kim}}]{Wang2022}%
  \BibitemOpen
  \bibfield  {author} {\bibinfo {author} {\bibfnamefont {P.}~\bibnamefont {Wang}}, \bibinfo {author} {\bibfnamefont {J.}~\bibnamefont {Zhang}}, \bibinfo {author} {\bibfnamefont {C.~Y.}\ \bibnamefont {Luan}}, \bibinfo {author} {\bibfnamefont {M.}~\bibnamefont {Um}}, \bibinfo {author} {\bibfnamefont {Y.}~\bibnamefont {Wang}}, \bibinfo {author} {\bibfnamefont {M.}~\bibnamefont {Qiao}}, \bibinfo {author} {\bibfnamefont {T.}~\bibnamefont {Xie}}, \bibinfo {author} {\bibfnamefont {J.~N.}\ \bibnamefont {Zhang}}, \bibinfo {author} {\bibfnamefont {A.}~\bibnamefont {Cabello}},\ and\ \bibinfo {author} {\bibfnamefont {K.}~\bibnamefont {Kim}},\ }\href {https://doi.org/10.1126/sciadv.abk1660} {\bibfield  {journal} {\bibinfo  {journal} {Science Advances}\ }\textbf {\bibinfo {volume} {8}},\ \bibinfo {pages} {1} (\bibinfo {year} {2022})}\BibitemShut {NoStop}%
\bibitem [{\citenamefont {Jerger}\ \emph {et~al.}(2016)\citenamefont {Jerger}, \citenamefont {Reshitnyk}, \citenamefont {Oppliger}, \citenamefont {Poto{\v{c}}nik}, \citenamefont {Mondal}, \citenamefont {Wallraff}, \citenamefont {Goodenough}, \citenamefont {Wehner}, \citenamefont {Juliusson}, \citenamefont {Langford},\ and\ \citenamefont {Fedorov}}]{Jerger2016}%
  \BibitemOpen
  \bibfield  {author} {\bibinfo {author} {\bibfnamefont {M.}~\bibnamefont {Jerger}}, \bibinfo {author} {\bibfnamefont {Y.}~\bibnamefont {Reshitnyk}}, \bibinfo {author} {\bibfnamefont {M.}~\bibnamefont {Oppliger}}, \bibinfo {author} {\bibfnamefont {A.}~\bibnamefont {Poto{\v{c}}nik}}, \bibinfo {author} {\bibfnamefont {M.}~\bibnamefont {Mondal}}, \bibinfo {author} {\bibfnamefont {A.}~\bibnamefont {Wallraff}}, \bibinfo {author} {\bibfnamefont {K.}~\bibnamefont {Goodenough}}, \bibinfo {author} {\bibfnamefont {S.}~\bibnamefont {Wehner}}, \bibinfo {author} {\bibfnamefont {K.}~\bibnamefont {Juliusson}}, \bibinfo {author} {\bibfnamefont {N.~K.}\ \bibnamefont {Langford}},\ and\ \bibinfo {author} {\bibfnamefont {A.}~\bibnamefont {Fedorov}},\ }\bibfield  {journal} {\bibinfo  {journal} {Nature Communications}\ }\textbf {\bibinfo {volume} {7}},\ \href {https://doi.org/10.1038/ncomms12930} {10.1038/ncomms12930} (\bibinfo {year} {2016})\BibitemShut {NoStop}%
\bibitem [{\citenamefont {George}\ \emph {et~al.}(2013)\citenamefont {George}, \citenamefont {Robledo}, \citenamefont {Maroney}, \citenamefont {Blok}, \citenamefont {Bernien}, \citenamefont {Markham}, \citenamefont {Twitchen}, \citenamefont {Morton}, \citenamefont {Briggs},\ and\ \citenamefont {Hanson}}]{George2013}%
  \BibitemOpen
  \bibfield  {author} {\bibinfo {author} {\bibfnamefont {R.~E.}\ \bibnamefont {George}}, \bibinfo {author} {\bibfnamefont {L.~M.}\ \bibnamefont {Robledo}}, \bibinfo {author} {\bibfnamefont {O.~J.}\ \bibnamefont {Maroney}}, \bibinfo {author} {\bibfnamefont {M.~S.}\ \bibnamefont {Blok}}, \bibinfo {author} {\bibfnamefont {H.}~\bibnamefont {Bernien}}, \bibinfo {author} {\bibfnamefont {M.~L.}\ \bibnamefont {Markham}}, \bibinfo {author} {\bibfnamefont {D.~J.}\ \bibnamefont {Twitchen}}, \bibinfo {author} {\bibfnamefont {J.~J.}\ \bibnamefont {Morton}}, \bibinfo {author} {\bibfnamefont {G.~A.~D.}\ \bibnamefont {Briggs}},\ and\ \bibinfo {author} {\bibfnamefont {R.}~\bibnamefont {Hanson}},\ }\href {https://doi.org/10.1073/pnas.1208374110} {\bibfield  {journal} {\bibinfo  {journal} {Proceedings of the National Academy of Sciences of the United States of America}\ }\textbf {\bibinfo {volume} {110}},\ \bibinfo {pages} {3777} (\bibinfo {year} {2013})}\BibitemShut {NoStop}%
\bibitem [{\citenamefont {Hofmann}(2023)}]{Hofmann2023}%
  \BibitemOpen
  \bibfield  {author} {\bibinfo {author} {\bibfnamefont {H.~F.}\ \bibnamefont {Hofmann}},\ }\href {https://doi.org/10.1364/opticaq.502468} {\bibfield  {journal} {\bibinfo  {journal} {Optica Quantum}\ }\textbf {\bibinfo {volume} {1}},\ \bibinfo {pages} {63} (\bibinfo {year} {2023})}\BibitemShut {NoStop}%
\bibitem [{\citenamefont {Elitzur}\ and\ \citenamefont {Vaidman}(1993)}]{Elitzur1993Bomb}%
  \BibitemOpen
  \bibfield  {author} {\bibinfo {author} {\bibfnamefont {A.~C.}\ \bibnamefont {Elitzur}}\ and\ \bibinfo {author} {\bibfnamefont {L.}~\bibnamefont {Vaidman}},\ }\href {https://doi.org/10.1007/BF00736012} {\bibfield  {journal} {\bibinfo  {journal} {Foundations of Physics}\ }\textbf {\bibinfo {volume} {23}},\ \bibinfo {pages} {987} (\bibinfo {year} {1993})}\BibitemShut {NoStop}%
\bibitem [{\citenamefont {Hance}\ \emph {et~al.}(2024)\citenamefont {Hance}, \citenamefont {Matsushita},\ and\ \citenamefont {Hofmann}}]{Hance2024}%
  \BibitemOpen
  \bibfield  {author} {\bibinfo {author} {\bibfnamefont {J.~R.}\ \bibnamefont {Hance}}, \bibinfo {author} {\bibfnamefont {T.}~\bibnamefont {Matsushita}},\ and\ \bibinfo {author} {\bibfnamefont {H.~F.}\ \bibnamefont {Hofmann}},\ }\href {https://doi.org/10.1088/2058-9565/ad63c7} {\bibfield  {journal} {\bibinfo  {journal} {Quantum Science and Technology}\ ,\ \bibinfo {pages} {1}} (\bibinfo {year} {2024})},\ \Eprint {https://arxiv.org/abs/2404.16477} {2404.16477} \BibitemShut {NoStop}%
\bibitem [{\citenamefont {Ji}\ and\ \citenamefont {Hofmann}(2024)}]{Ji2024quantitative}%
  \BibitemOpen
  \bibfield  {author} {\bibinfo {author} {\bibfnamefont {M.}~\bibnamefont {Ji}}\ and\ \bibinfo {author} {\bibfnamefont {H.~F.}\ \bibnamefont {Hofmann}},\ }\href {https://doi.org/10.22331/q-2024-02-14-1255} {\bibfield  {journal} {\bibinfo  {journal} {{Quantum}}\ }\textbf {\bibinfo {volume} {8}},\ \bibinfo {pages} {1255} (\bibinfo {year} {2024})}\BibitemShut {NoStop}%
\bibitem [{\citenamefont {Hofmann}(2024)}]{Hofmann2024}%
  \BibitemOpen
  \bibfield  {author} {\bibinfo {author} {\bibfnamefont {H.~F.}\ \bibnamefont {Hofmann}},\ }\href {https://doi.org/https://doi.org/10.3390/e26090725} {\bibfield  {journal} {\bibinfo  {journal} {Entropy}\ ,\ \bibinfo {pages} {1}} (\bibinfo {year} {2024})}\BibitemShut {NoStop}%
\bibitem [{\citenamefont {Silverstone}\ \emph {et~al.}(2014)\citenamefont {Silverstone}, \citenamefont {Bonneau}, \citenamefont {Ohira}, \citenamefont {Suzuki}, \citenamefont {Yoshida}, \citenamefont {Iizuka}, \citenamefont {Ezaki}, \citenamefont {Natarajan}, \citenamefont {Tanner}, \citenamefont {Hadfield}, \citenamefont {Zwiller}, \citenamefont {Marshall}, \citenamefont {Rarity}, \citenamefont {O'Brien},\ and\ \citenamefont {Thompson}}]{Silverstone2014}%
  \BibitemOpen
  \bibfield  {author} {\bibinfo {author} {\bibfnamefont {J.~W.}\ \bibnamefont {Silverstone}}, \bibinfo {author} {\bibfnamefont {D.}~\bibnamefont {Bonneau}}, \bibinfo {author} {\bibfnamefont {K.}~\bibnamefont {Ohira}}, \bibinfo {author} {\bibfnamefont {N.}~\bibnamefont {Suzuki}}, \bibinfo {author} {\bibfnamefont {H.}~\bibnamefont {Yoshida}}, \bibinfo {author} {\bibfnamefont {N.}~\bibnamefont {Iizuka}}, \bibinfo {author} {\bibfnamefont {M.}~\bibnamefont {Ezaki}}, \bibinfo {author} {\bibfnamefont {C.~M.}\ \bibnamefont {Natarajan}}, \bibinfo {author} {\bibfnamefont {M.~G.}\ \bibnamefont {Tanner}}, \bibinfo {author} {\bibfnamefont {R.~H.}\ \bibnamefont {Hadfield}}, \bibinfo {author} {\bibfnamefont {V.}~\bibnamefont {Zwiller}}, \bibinfo {author} {\bibfnamefont {G.~D.}\ \bibnamefont {Marshall}}, \bibinfo {author} {\bibfnamefont {J.~G.}\ \bibnamefont {Rarity}}, \bibinfo {author} {\bibfnamefont {J.~L.}\ \bibnamefont {O'Brien}},\ and\ \bibinfo {author} {\bibfnamefont {M.~G.}\ \bibnamefont {Thompson}},\ }\href
  {https://doi.org/10.1038/nphoton.2013.339} {\bibfield  {journal} {\bibinfo  {journal} {Nature Photonics}\ }\textbf {\bibinfo {volume} {8}},\ \bibinfo {pages} {104} (\bibinfo {year} {2014})}\BibitemShut {NoStop}%
\end{thebibliography}
%

\end{document}